\documentstyle[preprint,revtex,eqsecnum]{aps}
\begin{document}
\draft
\newcommand{\be}{\begin{equation}}
\newcommand{\ee}{\end{equation}}
\newcommand{\bi}{\begin{instit}}
\newcommand{\ei}{\end{instit}}
\begin{center}
 {\Large \bf A Model for Fracture\\ in Fibrous  Materials}
\end{center}
\author{A. T. Bernardes }
\bi
Departamento de F\'{\i}sica - ICEB \\
Universidade Federal de Ouro Preto \\
Campus do Morro do Cruzeiro \\
35410-000 Ouro Preto MG - Brazil
\ei
\author
{J. G. Moreira }
\bi
Departamento de F\'{\i}sica -Instituto de Ci\^encias Exatas \\
Universidade Federal de Minas Gerais \\
C. P. 702, 30161-970 Belo Horizonte MG - Brazil
\ei
\newpage
\begin{center}
ABSTRACT
\end{center}
\begin{abstract}
 A fiber bundle model in $(1+1)$-dimensions for the
  breaking of fibrous composite matrix is introduced. The model
consists of $N$ parallel fibers fixed  in two plates.
 When one of the plates is pulled in the direction parallel to the fibers,
 these can be broken with
 a probability that depends on their elastic energy.
The mechanism of rupture  is simulated by the breaking of
neighbouring fibers that can generate
random cracks  spreading up through the system.
 Due to the simplicity of the model we have virtually
no computational limitation.
The model is sensitive to external conditions as temperature and traction
time-rate.  The energy {\it vs.}
 temperature behaviour, the diagrams of stress {\it vs.}
 strain and the histograms of the frequency {\it vs.} size of cracks
 are obtained.
\end{abstract}
\section{Introduction}

Fracture is an important problem in material sciences and
engineering. The response of a solid under load depends on the
 features of the material, the external conditions
 (temperature, humidity etc) and how
the load is applied (uniaxial, radial, shear etc). The main features
of the fracture processes can be found in the classical Young's experiment.
Let us consider a homogeneous bar of initial length $L_o$ and cross section $S$
pulled  by an uniaxial force $F$ parallel to the length. In the
$\sigma =F(t)/S(t)$ {\it vs.} $\delta =\Delta L/L_o$ diagram one can observe
 an elastic (linear and nonlinear) region and a plastic/deformation one.
 The elastic region occurs in the   beginning of the
traction when the material returns to $L_o$ if the traction is stopped.
 On the other hand, the material acquires a permanent deformation when the
 force vanishes in the plastic/deformation region.
 If the material breaks in the elastic regime the fracture
is called brittle (like glass at room temperature). Otherwise, if
the material breaks in the plastic/deformation region the fracture is
called ductile (like a school-rubber).

The presence of disorder in the material  is an important feature
that determines  the rupture processes \cite{hansroux}.
 These inhomogeneities strongly
  influence the mechanical behaviour of the material and are
responsible for the patterns obtained experimentally.
 In the last decade, some models taking into account this feature were proposed
 to simulate the breaking processes of disordered media \cite{hans89}.
  The material, in general, is represented by a network of structural
 units whose rate of rupture depends on the local conditions and
inhomogeneities.
These models, which were proposed to simulate the rupture of polymer fibers or
thin films (models of lattice springs) \cite{termonia85ms,termonia86m,meakin87}
and to study the interface
properties of breaking processes \cite{lucila85rh,lucila89hhr},
 have been studied mostly by computational experiments .
  However, these models provide just a partial description
 of the problem. At most, only the fracture pattern and the stress
{\it vs.} strain  diagram can be obtained. These models  do not allow an
analysis of the dependence of rupture features with traction
velocity and temperature because they are sensitive to
changes  only in one of the external conditions.

In this paper  a fiber bundle model
to simulate the failure processes of fibrous material is introduced.
Fracture of fiber-reinforced materials is an important field
of investigation, because these materials have a higher Young's
modulus  and other different mechanical
properties than unreinforced ones \cite{llorca,grove93}.
Fiber bundle models were introduced to study the strength of material
where fibers are held together by friction forces. They are also used  to
study the breaking of composite materials where the fibers of the material
are joined together by other homogeneous material, as fiberglass-reinforced
 composite. When a fiber fails, the load that it carries is shared by
 intact fibers in the bundle. An important effort to
 study  these models was carried out by the calculation of
 the cumulative breaking probability of the chain of fiber bundles
\cite{harlow91,smith80,smith83}.

Our model considers the amount of elastic energy into the material,
the spread of a local crack and the fusion of cracks as the breaking
mechanism. Some features already proposed in the literature are used in the
definition of our model -- the computation of breaking probability from the
elastic energy of a fiber \cite{termonia85ms,termonia86m,meakin87}
 and a deformation limit for
an isolated fiber like the threshold in the random fuse network model
\cite{lucila85rh,lucila89hhr}. In addition, we adopt the cascade of breaking
fibers as the mechanism to form the cracks into the fiber bundle. This last
characteristic is clearly inspired in the self-organizing criticality
\cite{bak88}.
Our attention is focused on computational simulation
for the breaking of a fiber bundle when we have an uniaxial force
 (parallel to the fibers) in  $(1+1)$ dimensions. The fracture processes are
described by the  energy of the
rupture process {\it vs.} temperature, the diagram stress {\it vs.} strain
 and by the size of the cracks that occurs in the breaking. This paper
is organized as follow: in section II, the model
 is presented; the results of the computational experiments are shown and
discussed in section III; finally, the conclusions are given
 in the last section.

\section{The Model}

Our model  consists of $N_o$ parallel fibers, each of them with the same
elastic
constant $k$. These fibers are fixed in parallel plates as is shown in Figure
1.
  Note that  the first and
last fiber contact  with only one neighbour while the inner fibers have two
neighbours. For convenience one plate is fixed
 and the other is pulled by a force {F} in the direction parallel to the fibers
with constant velocity $v$. It means that at each time step $\tau$ the
amount of deformation of the non broken fibers is equal
 to ($\Delta z =v\times \tau$), where $v$ is the velocity (in our units
$\tau=1$).
When the deformation is $z$, the elastic energy for each fiber is given by
\be
\epsilon = {1 \over 2} k~z^2 ~~.
\ee
\noindent We define the critical elastic energy for each fiber as
\be
\epsilon_c=  {1 \over 2} k~z_c^2~~,
\ee
\noindent where $z_c$ is imposed as the maximum deformation
supported by an individual fiber.
We assume that an isolated fiber has a purely linear
elastic behaviour with a breaking probability which grows up with
the deformation $z$ of the fiber, being equal to unity at
$z=z_c$. The probability of rupture of the fiber $i$ is
\be
P_i (z) = {1 \over {(n_i+1)}} \exp[{1 \over t} (\delta^2~-~1)]~~.
\ee
\noindent Here $n_i$ is the number of non broken neighbours fibers
of the fiber $i$ (in this paper $n_i$ could be 0, 1 or 2),
\be
t ={k_B T \over \epsilon_c}
\ee
\noindent is the normalized temperature, $k_B$  is the Boltzmann constant
(in our unity system  it is equal 1) and
\be
\delta ={ z \over z_c}
\ee
\noindent is the strain of the material.
The dependence on the non broken neighbours fibers simulates
 the existence of an interaction between the fibers.
This dependence  is responsible for the  distribution of the load between
neighbouring fibers and allows a fiber having an elastic energy greater than
$\epsilon_c$. In this sense, one can observe fibers with $z > z_c$
if they have at least one  non broken neighbouring fiber.

 Initially all the fibers have the same length and zero deformation.
 In each time step of the simulation the system is pulled
 by  $\Delta z$, and we  randomly choose $N_q= q \times N_o$ fibers that
 can be broken, where $q$ is a positive number.
 It means that the  probability  of rupture for
 the material does not depend on the number of fibers in the fiber bundle.
 This assumption is in agreement with the observation that
 systems with different sizes must have the same rupture features for the
 same external conditions (temperature and traction velocity).
 Obviously the force and the energy
 needed to break the bundle must depend on the system size
 but not the stress {\it vs.} strain diagrams or the
size of the cracks that arises in the breaking processes.
This assumption makes also possible the appearance of cracks
 in different parts of the material for the same deformation.
Let us consider a chosen fiber. The breaking probability is
evaluated and compared
 with a random number in the interval $[0,1)$.
If the random number is less than the
breaking probability, the fiber breaks. The load
spreads to the neighbour fibers and
 the breaking  probability of them increases because of
the decreasing of the parameters $n_{i-1}$ and $n_{i+1}$.
This procedure describes the propagation of the crack through
 the fiber bundle. Then, the same steps are done for
one of the neighbouring fibers. Note that if it breaks, a cascade begins.
It stops in a given fiber,
when the test of the probability does not allow its rupture,
 or when  a hole in the bundle  is found (an old crack).
  The propagation of the crack is done in either ``left'' or ``right''
directions,
 perpendicular to the force applied on the system.
 When the cascade process stops, other
 fiber in the $N_q$ set is chosen and all steps already described are repeated.
After the  $N_q$ trials, we pull the system to
 a new displacement $\Delta z$ and the breaking procedure begins again.
The simulation continues until  the rupture of the system,
when no more entire fibers exists.

\section{Results and Discussions}

At $t=0$ it is easy to see that the model breaks at $\delta=1.0$ with a maximum
force $F=N_o k z_c$. All the fibers break at the same time and we
have just one crack spreading in the entire system
(the limit of a brittle fracture). For finite temperatures  different
behaviours are observed when the traction velocity is varied.
The number of fibers is chosen in such a way that it does not affect the
propagation of the cracks. It means that a crack greater than
or equal to the size of the
system, for the values of $t$ and $v$  used in the simulation, occurs with
a negligible probability.
In order to investigate this picture we have performed simulations
in systems $N_o=10^3 - 10^6$. This probability
is controlled by determining the distribution of cracks {\it vs.} the sizes
of the cracks arising in the process of fracture. We have used
the following values for the parameters: $q=0.1$, $N_o=10^4$, $z_c=1$ and
$k=1$.

Preliminary  we have obtained the stress {\it vs.}
strain diagrams for different temperatures and traction velocities.
  When the deformation of the bundle  is $z$ and
the number of non broken fibers is $N$, the stress $\sigma$ is defined as
\be
\sigma ={Nkz \over N_o}~~.
\ee
\noindent  The strain $\delta$ was defined in expression (2.5).
 We compare our results with the description
obtained experimentally in order to classify the fracture
as  brittle or ductile \cite{gordon93}.
 Figure 2  shows the result of a computational simulation carried out
in just one fiber bundle. In this case, averages are avoiding. For $t=0.1$ one
observes a brittle behaviour, i. e., the fiber bundle breaks in the
 elastic region. Note that the $\sigma \times \delta$ plot is purely linear for
the highest velocity ($v=0.1$).
At  $t=1.0$ and for high and intermediate velocities the fracture occurs
in the brittle/ductile transition region.
 The rupture of the material is ductile  for low velocities and
it occurs in the plastic/deformation region. For high temperatures ($t=4.0$),
the shape of the stress {\it vs.} strain plot is typically
ductile for intermediate and low velocities. For high velocities
 the fracture occurs in the transition region brittle/ductile.

Now let us discuss the behaviour of the energy of rupture as function of
temperature. This energy is defined as the work done to break the material
 and it can be obtained from the stress {\it vs.} strain diagrams.
It is well known that the breaking of materials has  strong dependence with
temperature. In general, some materials break brittle at low temperature
and ductile at high ones. It means that the energy of the fracture process has
a
small value in the brittle region and a greater value in the ductile one.
The results of the normalized averaged energy  of the breaking process
{\it per} fiber $<E_f>$ {\it vs.} the normalized temperature $t$ are shown in
 Figure 3. We have considered $10^3$ samples with $10^4$ fibers,
 with velocities $v = 0.001,~0.002$ and $0.005$ in the simulations. At low
 temperatures
the energy of the fracture becomes independent of the traction velocity.
 For velocity $v=0.005$ the energy increases with temperature. On the other
hand, for slow traction ($v=0.001$) the energy grows up to a maximum
 (near $t\sim 0.5$) and for $t>0.5$ it decays smoothly. For an intermediate
value of the velocity ($v=0.002$), the energy remains closely
constant at high temperatures.

Figure 4 shows the frequency of the cracks $H_c$ {\it vs.} the size of the
cracks $S_c$ that arises in the breaking process. The frequency of the cracks
is averaged over the samples ($10^3$ ones in this simulation). Two
features can be observed in this figure. For a low temperatures ($t=0.1$,
typically brittle fracture) one observes cracks of very different sizes.
For low velocities one observe cracks with a maximum size $\sim 10^2$.
For high velocities ($v\approx 0.1$) the size of the cracks tends to the
 entire system ($10^4$ fibers). This means that the system was
pulled essentially non broken until a certain time when a big crack arises
 in the material. After this big crack, small ones are observed because
of the rupture of the remained fibers. The brittle process is characterized
 by the existence of
cracks with different sizes and a remarkable feature is the presence of cracks
with sizes near to the system size. The curves that represent $H_c$
{\it vs.} $S_c$ have  a maximum at $S_c = 1$. Note that at the
 beginning $H_c$ goes down linearly.
In order to verify this last feature, we  adjust the data using
\be
H_c \sim S_c^{\alpha}~~.
\ee
 A good fit for this
linear part is obtained with $\alpha \sim 1.02$ (see Figure 5).

As long as the temperature is increased the size of the cracks becomes
 smaller.
When one has many cracks of small sizes the fracture is clearly ductile.
These cracks appear in different parts of the material and the shape of the
curve $H_c$ {\it vs.}  $S_c$ changes. The curve has a maximum at $S_c=2$,
instead $S_c=1$ that occurs for brittle fractures.

A quite different feature can be observed for a slow traction ($v=0.001$)
and high temperatures $(t>0.5)$ when we have obtained a ductile fracture with
an unusual low energy of rupture (see Figure 3).  Only cracks of small
size  are present in the system $(S_c<12)$. Now the system fails when
  a force smaller than the one needed to break it at higher velocities is
applied. This could indicate
that the system is in a different state at high temperatures and
that we have observed it in disaggregation.

\section{Conclusion}

We have introduced a fiber bundle model for simulate fractures in fibrous
materials. The model is sensitive to external conditions which are
present in some problems of material sciences: traction velocity and
temperature. The simplicity of the model allows us to perform
computations on very large systems. Because of this feature we
can explore all pictures of the failure processes.

We have obtained stress {\it vs.} strain  diagrams showing
features of the two principal types of fractures: brittle and ductile.
For low temperatures the system breaks brittle, independent of the
traction velocity. When the temperature increases, the fracture is
influenced by the traction velocity. We can observe a transition
from  the brittle regime to  the ductile one. The amount of energy needed to
rupture the material is dependent on the traction velocity. For high
velocities  more energy is needed. This comes from the fact that the size
of the cracks depends on the temperature. For high temperatures
and low velocities we observe a curious behaviour.
In this case the energy is smaller than that needed to break
the material in brittle regime. This could indicate that we have a
disaggregation process at this temperature and
the interaction between the fibers exerts a small
influence in the rupture process.
These results are independent on the number of fibers in the fiber bundle,
because we have chosen values of the parameters $v$ and $t$ for which
the maximum crack sizes obtained in our simulations are less than
$N_o$.

 Several questions remain opened. The first one is the behaviour of this model
in $(2+1)$-dimensions for several lattices topologies.
 This could  allow  the comparison of our results with
that observed in realistic systems. The behaviour of the model in high
temperatures and low velocities needs a more accurate investigation.
It is interesting to verify if those features remains in $(2+1)$-dimensions.

We thank J. Kamphorst Leal da Silva, J. A. Plascak
and B. V. da Costa  for helpful criticism of the manuscript.
One of us (ATB) acknowledge
the kind hospitality of the Depto de F\'{\i}sica, UFMG.

\bibliographystyle{revtex2}
\bibliography{refer}
\newpage
{\large \bf Figure captions}

{\bf Figure 1:} A schematic representation of our model. We have filled
in deep gray the fibers fixed  on a plate at rest (bellow) with the
high extremity fixed on a moving plate, pulled with constant displacement.

{\bf Figure 2:} Strees {\it vs.} strain plots for different normalized
temperatures
(indicated in the diagrams) and velocities ($v=0.001$ - full line; $v=0.01$
- dashed line; $v=0.1$ - long-dashed line).
The pictures were made with just one simulation with a
sample of $10^4$ fibers for each pair of parameters.

{\bf Figure 3:} Energy of fracture process {\it per} fiber $<E_f>$ {\it vs.}
normalized temperature $t$. The value of the velocities for each curve are
shown in the inner box.

{\bf Figure 4:} Frequency of cracks $H_c$ {\it vs.} crack sizes $S_c$
 for different temperatures and velocities.
The simulations were performed in $10^3$ samples of $10^4$ fibers for
each pair of parameters.
Pictures at the top have been calculated for $t=0.1$, at the middle
 for $t=1.0$ and at the bottom for $t=4.0$. Full line represents $v=0.001$,
 dashed line $v=0.01$ and long-dashed line $v=0.1$.

{\bf Figure 5:} Regression (dotted lines) for the diagrams frequency
of cracks $H_c$ {\it vs.} crack sizes $S_c$ for $t=0.1$ for different
velocities (indicated in the inner box).
\end{document}